# COMBINING AGILE WITH TRADITIONAL V MODEL FOR ENHANCEMENT OF MATURITY IN SOFTWARE DEVELOPMENT

**Ahmed Mateen[*]**

**Madiha Tabassum[*]**

**Akmal Rehan[*]**

**Abstract**

In the field of software engineering there are many new archetypes are introducing day to day Improve the efficiency and effectiveness of software development. Due to dynamic environment organizations are frequently exchanging their software constraint to meet their objectives. The propose research is a new approach by integrating the traditional V model and agile methodology to combining the strength of these models while minimizing their individual weakness.The fluctuating requirements of emerging a carried software system and accumulative cost of operational software are imposing researchers and experts to determine innovative and superior means for emerging software application at slight business or at enterprise level are viewing for. Agile methodology has its own benefits but there are deficiency several of the features of traditional software development methodologies that are essential for success. That's why an embedded approach will be the right answer for software industry rather than a pure agile approach. This research shows how agile embedded traditional can play a vital role in development of software. A survey conducted to find the impact of this approach in industry. Both qualitative and quantitative analysis performed.

**Keywords:** Agile software development;Traditional V model;Software quality;Integration;Software maturity.

[*] **Computer Science Department, Agriculture University Faisalabad, Pakistan**

# 1. Introduction

Whenever work start on software development project, there exist certain complications that are it goes in excess of schedule? Pass on over budget? Didn't meet the customer requirement? In fact software development is an extremely vigorous procedure so by locating the exact things to the actual place at the exact stage is the core principle to solve the difficulties.

There are the different kind and different intellectual level of stakeholders that have different responsiveness, activities and views with respect to the requirements. For this agile embedded traditional V model approach is best fit. It's a user centered approach. There would be lowest chances of software failure if all indicated stakeholders confirm completely known requirements. It's a most problematic task to visibly detect the requirements for altered deliverables as it prove the software that it will be effective or not. It became more problematic to select for purely dispersed stakeholders for requirements [1].

For presenting innovative and fresh software development technique, it is assuming that it will enhance the efficiency and productivity of software designers, testers, in short entire software crew. Software production can be explained as proportion of productivity manufactured by software team and assets used by them. The authorities formulated productivity questions are: Is software team performance greatest as it consider? Are team participants in opposition? Is effort being successful? Hence entirely replied gather by applying agile embedded traditional V model. By applying this approach for software development and testing team will be able to produce extremely observed documents to determine assets that are utilized to calculate the sum of productivity produce. For the assessment of software dimension between development period features of agile embedded v model will improve software team output. Organizations are using agile due to its flexibility of time period epoch and plan.

Agile development methodology give emphasize on people instead of processes. Maturity enhance by less attention on detail processes are define and enhance perception of agile practitioners and quantity of practices with content investigation of the qualitative data [2]. Maturity in agile enhance maintenance, assurance, cooperation, and proficiencies in organizations among development team [3]. When stockholders does not understand what they want it becomes difficult to develop a desired product as the user want or if not properly understand it may lead to failure of developed software.

This feature of agile are flexible lead to various problem of integration. Whereas entity modules can be developed completely useful cannot function as an application when integrated [4]. Methodologies agile software development characterized by small cycle iterative and incremental development user participation in improvement time box mutual decision making merger of rapid response. When there is confidence in people instead of predefined process similar to a socio technical procedure of the team members, client, company an interaction designer and a member of the other team members [5]

1.1. Agile software development

Mostly agile technique break the assignments into little augmentations with insignificant arranging and don't straightforwardly include long haul arranging. Arranging, prerequisites investigation, outline, coding, unit testing, and acknowledgment testing.

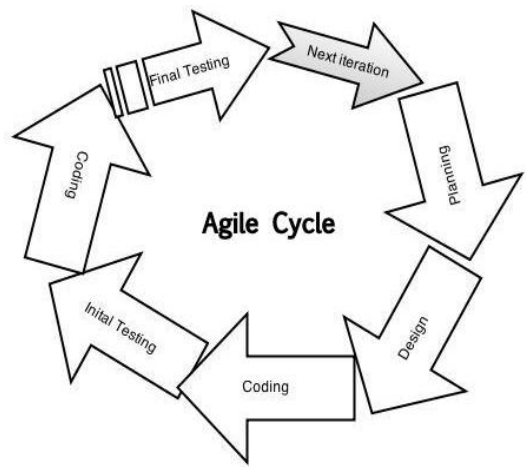

Figure 1. *Agile software development cycle*

1.2. Agile principles

The Agile Manifesto depends on standards.

- Customer fulfillment by ahead of schedule and consistent conveyance of important programming
- Welcome evolving necessities, even in late improvement
- Close, every day participation between businessmen and engineers
- Projects are worked around persuaded people, who ought to be trusted
- Face-to-face discussion is the best type of correspondence.

**2. Traditional V Model**

A programming procedure model is a theoretical representation to outline the procedure from a specific planned. There are number of general model of programming procedure, for example, water fall, transformative advancement, formal framework improvement and reuse base improvement, and so forth. The customary procedure model comprises of grouping of stages which implies that before entering to the following stage the earlier should be finished. Toward the end of this stage prerequisite detail archive is acquired. That is a marked archive from both side designer and customer. This prompts the following stage that is framework plan. After advancement each unit is tried for its usefulness which is alluded to as unit testing. After joining the entire framework is tried for any shortcoming any disappointment. Whenever useful and non-practical testing is performed then item is sent or discharged in nature. At operational stage a few issues might be emerges.

2.1.V model relation to agile software development

Agile has much wide advantages over plan driven development. V model and standards apply in programming improvement. They give answers for the issues when they use in programming advancement. Agile as of now has enhanced the generation quality rather than arrangement driven methodologies.

- Continuously cooperation with client to include appreciation.
- Customer joint effort all through the development of software.
- Requirements changing are welcomed.
- Throughout development of software to fulfill the client.
- Throughout the requirement changing are invited.

## 3. Characteristics of the Purposed Approach

3.1. Reduction of project risk

The reason advancement model enhances the effortlessness of the task and venture control by determining reliable methodologies and unfurling the subsequent results and responsible parts.

3.2. Improved quality

As the procedure model uses set of institutionalized procedures that guarantee the discoveries to be given are finished and having wanted quality.

3.3. Total cost reduced

By selecting fitting model for every improvement stage expense of the venture diminished.

3.4. Scalability

It is the capacity of the half breed process model to be versatile in term of venture size. It functions admirably for little to mega estimated ventures

3.5. Management and development of model

The administrators and group of the product engineers have their own obligations that are appointed before the incitation of the project.

- Requirements are arranged into sets by breaking down them.
- Assign the specific tasks among development team.
- Development group communication with the clients to get the point by point requirements.
- Deliver the complete undertaking with its documentation to mangers.

## 4. Review of Litrature

Various agile approaches near about fifty-eight practices applied to check their effectiveness. Almost fifteen factors were study and associated with their practices success of the project. These factors incorporate with the improvement practice of agile process. They applied Correlations among the mined features were intended for exploration to the resulting matrix and significant correlation outcomes recommended for iterative and incremental development.

Different well known fifteen factors enterprise forms invalid analysis lightweight testing architecture and arrangement traditional prominence affirmation program ethics insignificant requirements iterative and incremental development and communication process database training communication exercise and agile practices assertion communication team program consideration and review adaptation were analyzed [6].

Software is a gathering of programs and related facts and figures that provides the guidelines to articulate a computer what to do and how to do this and how it could be performed [7]. The collection of programs and information construct the structure of software. Any update is called a change or modification in structure of software. Dynamic changes occur in a system without any disturbance upgrades any running system is a type of software development.

The Agility Profiler a matrix that initiatives struggle for occasionally radical renovations to stand and expand their effectiveness. Below present unclear and unsettled situations in several competitive situations agility is frequently required and considered a feasible strategy. Agility requirements are well known in what way its existing competencies can be enhanced and difficulties can be minimized [8]. The profiler supports company fined motorist aims resources and obstacles to refining agility. There is no that type of thing which are bad or good a profile as a profile it specifies the strategy of organizations.

In agile software development restricted consideration is focused to starting the hypothetical foundations. Consequently, between the chances and trials that lie ahead for investigators is to differentiate inconsistency across the experimental works [9]. V model identifies for its altered challenging types like modular testing and integration testing. There exist associations among Confirmation and Authentication stages in development process. If there is presence of several errors through development due to incompletion of various phases then on opposite adjacent of the resultant platform is repeated again [10]. The reason for choosing the v model during medical device software development is fit with inflexible requirements as it guides the institute to create necessary releases of software to test the authoritarian conformance.

The impacts of agile approaches in business developments and try to determine in what type of conditions this is valuable to put on such approaches [11]. Although the major focus becomes on the evolvement of the agile association with their achievement. The major outcomes observed for application of agile approaches were applicable the capability to priorities for management of changes and for basic development procedure quality, user fulfillment and satisfaction for project reflectivity.

Maturity has border definition and consider that Agile Maturity Model could become helpful useful for the organizations. Maturity addressed with the framework specified feature of teams. That checks the applied applicability of the outcomes accessible there [12]. Additionally, the applied applicability of outcomes is validated in fresh readings that consider Agile Maturity Models approving the requirement for fresh innovative methodology to increase maturity in agile. By amending v model in dividing method, traceability can be certain of effort portions. Through that method constraint associated effort portion is belonging to minimum single project item and individual test category work article [13]. Design associated work slice to smallest than one component and test category item. At least segment sort work item is linked to single test category of item.

Traditional programming advancement techniques are rigid and neglect to react on forceful client demands. Interestingly, light-footed programming philosophies give an arrangement of practices that take into account speedy adjustments coordinating the current item improvement needs[14] In spite of the fact that the estimation of the spry philosophies is well demonstrated for agile. Upgrades are seen on the quality and on the client impression of the finished item, while agile techniques take into account prerequisite changes even late in the venture. In the meantime, constructing better correspondence and cooperation in the group as a result of taking after the agile practices, results to upgraded relations between colleagues and to enhanced representative fulfillment measurements.

For adherence to development levels by utilizing lightweight procedures that require low levels of exertion is viewed as a test for programming advancement associations. This study tries to assess, orchestrate, and exhibit results on the utilization of the Capability Maturity Model Integration (CMMI) in mix with lithe programming improvement, and from that point to give a diagram of the subjects looked into, which incorporates a talk of their advantages and impediments, the quality of the discoveries. The strategy connected was a Systematic Literature Review on studies distributed. The pursuit methodology distinguished 3193 results, of which 81 included studies on the utilization of CMMI together with agile philosophies. The advantages found were gathered into two principle classifications [15] those identified with the association when all is said in done and those identified with the improvement procedure, and were composed into subcategories, as per the territory to which they allude.

## 5. Methodology

A survey conducted to check the consequences of new approach developed by mixing up the agile and v model. The effect of this integrated approach checked through a survey on software industry. Different attributes are checked and its effects on software on development. The survey is done to ensure what will be the effects on software industry software maturity Attribute, while using standards for safety critical software systems.

Software maturity achievement is the basic purpose of this research. According to customer quality is to fulfill his requirement at an acceptable level of cost. According to user quality is easy to use and work with software. According to developer quality is easy to design, maintain and reuse software's parts and according to developer quality is low development cost as well as high customer satisfaction. To fulfill these demands our approach will use a questionnaire that filled by different software persons related to software industry.

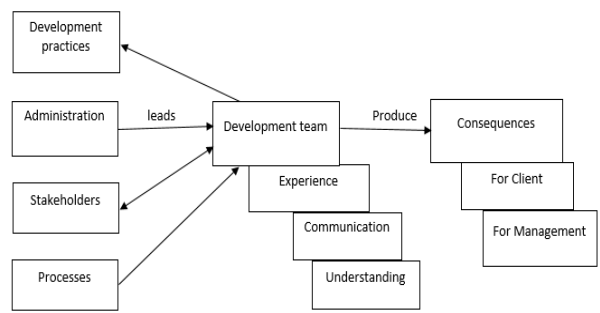

Figure 2:*Perceptions of maturity in agile software development*

5.1.Agile integrated traditional Development

Agile incremental models likemethods help to provide feedback concluded through well communication a due to slight iterations and client collaboration. The transferal of knowledge, the advantage of communication supports additionally. To have the client agile approaches promote plan. This is observed as appreciated by developers as they be able to get control above practices frequent opinion by the clients as this being offers appreciate projects. Additional advantage providing to clients is the consistent opinion on development. Agile developments are perceived as effort comfortable situation assistance maintaining value of working life span that can be considered as trustful, and In agile methodology well-known complications are that architecture not have sufficient emphasis and agile methodology does not scale well. Constant testing and combination is a main idea [16]. Various platforms and method dependences however, understanding constant testing needs great work as generating an integrated check situation is tough for. Testing is a bottleneck for security basic frameworks Moreover, in agile developments, testing done very frequently because security of critical system the purpose is that and at the same time carefully developed. Team associates have to be vastly practiced on the level of team. There are few benefits have been point out by the regard to the customers site. Customers have to obligate sets them below stress the problem that they obliges their obligation done a lengthy period of time for the entire development procedure. The basis of the info was the survey and for agile the paper systematic review of literature focused for full analysis. Agile methodology leads to benefits the central outcome was the development processes. Research framework for agile methods characteristics Moreover the need has been distinguished to describe the setting of the procedures contemplated.

5.2 RS Stage:

o       RS means Requirement specification, the things those have been noticed in this phase of development cycle are:

–       To find out the detailed requirement.

–       Requirement in design and implementation phase.

–       To gather the requirement in testing phase.

o       Prioritization of requirement that which requirement is necessary to implement first.

o       Scheduling the requirements according to their priority that which requirement is need to implement after which one.

5.3 DD stage:

o       DD means design and development stage, in this step of development cycle the important point to remember which will be implement are:

o       Low level designing of project to be desire.

o       Development of all requirements low and as well as high.

5.4 TA stage:

o       TA means Testing and acceptance stage, in this step of development cycle the important point to remember are the following:

o       Design of test Cases (There will be one test case for one Individual Requirement, every requirement will have its own test case design for specifically to check that requirement and with its implementation checking.)

–       Unit Test

–       Acceptance Test

–       Integration Test

–       System Test

–       Function Test

Where unit testing outcomes only like binary either pass or fail means yes or no. Like unit testing the acceptance test suggests also what is actually the reason of failure and applicable things that can be also tell. Integration is also called definition of done, testing in which both components (software and hardware) are combined to assure either that they both can interact and cooperate with each other according to expectations and requirements to give the user the required output. There should be a testing environment exactly so close reflects to original.

5.5.DR stage

o       DR means Developed and Released stage, in this step of development cycle the product being developed fully after iterations and released to the customer provided to customer after validation and verification to the customer.

## 6. Result and Discussion

To carry out this research 20 questions were made by analyzing the previous history of agile and traditional model to check the quality attributes as design specification and requirement specification. Different software houses were visited and they answered the survey question. The respondents of the survey included as under: software systems engineer, IT managers, software engineers and software developers. The answers collected from the different background having respondents, they provide different views and analysis and in what way reuse includes practical and non-technical points of view. About 50% of the defendants (28 contributors) appealed to attain achievement in software schemes owed to reuse of the software which is a great advantage.

89.8% male and 10.2% female respondents encountered from total 50 respondents. Type of engineers were mostly software engineers and 98 % responses about their experience.

6.1. The Effect of moving towards agile integrated traditional Software Development Approach

The preferred tasks, properties of software all plan-driven methodologies segment some features those essential to be definite from the earlier until the completion of the development a complete strategy is erected modification request procedure is carry out and requirements are indicated in extraordinary way the whole design implementation start the programming work is only concentrated in the programming segment. Testing is done at the end of the project and the quality assurance is controlled in the formal method. For plan driven approaches as a representative it has been identified in empirical research that waterfall has to face challenges in development and there are many issues for the waterfall failures approach. Management of a large scope is the main factor identified, for example if the requirements cannot be managed well that must be the main reason for failure. The customers' at the end of the project current needs have been not addressed, results in that many features of the system were not implemented. Simply minor portion of developed code has been deployed and shown by the study of 400 waterfall projects. Lack of opportunity and change of needs are the reasons for this to clarify the misunderstandings. Customer's lack of opportunity caused to provide feedback on the system. The study was conducted in small software companies. The study identified related to the use of rational unified process negative and positive factors. The clear definition of roles and responsibilities of everyone the supportive process are the benefits of it. There is another model that is v model by directing on the v model it has not been capable to detect industrialized case studies though comparing different process models it was part of an experiment. In practice Plan-driven approaches are still widely used in many research articles as recognized.

Table 1. integration of both feature agile and traditional

|       |       | Frequency | Percent | Valid percent | Cumulative percent |
|-------|-------|-----------|---------|---------------|--------------------|
| Valid | yes   | 25        | 51.0    | 51.0          | 51.0               |
|       | no    | 24        | 49.0    | 49.0          | 100.0              |
|       | Total | 49        | 100.0   | 100.0         |                    |

Table 1 shows the effects of integration of both agile and v model and also showing the results of survey question which was asked from respondents. The respondent's answer of this question in yes or no and this table shows the calculations of that with respect to respondents' answers.

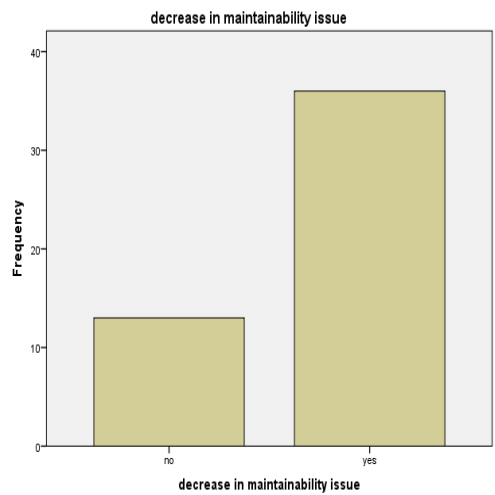

Figure 3.*Decrease in maintenance issues*

Fig 3 describe issues relate to the maintainability of software which arise different issues in development by applying the proposed approach will prove effective in handling maintainability issues. The majority of respondents gives positive response with respect to this question.

6.2.To check the effects by combining

Question was that what check the effects of combining agile and traditional model the focus of developers was combination of both during development of software. The following info collected show in table.

Table 2. effect on development process by combining

|  |  | Frequency | Percent | Valid Percent | Cumulative percent |
|---|---|---|---|---|---|
| Valid | no | 13 | 26.5 | 26.5 | 26.5 |
|  | yes | 36 | 73.5 | 73.5 | 100.0 |
|  | Total | 49 | 100.0 | 100.0 |  |

Table 2 explain the effects of development if applied the proposed features of agile and traditional with respect to survey results.QA carries the options of this question in the form of yes and no on the basis of the survey calculation show in table.

Table 3. improved quality by implementation of model

|  |  | Frequency | Percent | Valid Percent | Cumulative Percent |
|---|---|---|---|---|---|
| Valid | no | 12 | 24.5 | 24.5 | 24.5 |
|  | yes | 37 | 75.5 | 75.5 | 100.0 |
|  | Total | 49 | 100.0 | 100.0 |  |

Table 3 describe enhancement of quality in specification with respect to design, performance how get the quality or can enhance or not by combining agile and traditional method. The survey results are significance because due to changing of requirement traditional model not adopt modification in requirement in such agile development is best practice through in such way this agile embedded traditional will prove beneficial.

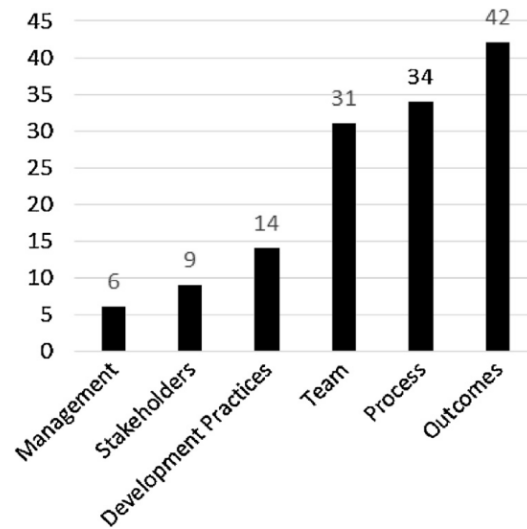

Figure 4.*Occurrence of Frequency*

Fig describe that the frequency of occurrence of all the involved entities how all these entities effecting the outcomes of the resultant product and at which level performing their role for outcomes. The outcomes of the resultant product depend upon all these occurring frequencies. Enhancement of maturity depend upon have an experienced development team.

6.3.Usage of agile integrating with v model software development Results

Usage of agile software development approach in the respondents' organizational unit was reported by 65%. The usage of agile software development methods would be help said by 65% because they agreed on about more than 50% benefits of agile. And left were not agreed or not sure and those were about 35%.

The benefits those were accepted by the respondents below there is number of respondents who were accepted that quality attributes about lean. The graph also displayed which shows their level of acceptance.

## 7. Conclusion

In this paper a survey conducted to check the result of integration of both agile and traditional v model and systematic review of literature conducted to check out the results of this integrated approach. This is an empirical study for software industry. The results of this survey provides positive responses for better enhancement of quality, efficiency, suitability, maintainability, and reduces the time and cost which is a great benefit. There are some limitations in this research however it is suggested that for further research to enhance this research as in this work the sample size is very low and data collected is small. If the size of data is small it will help in understandings more effectively and clearly.

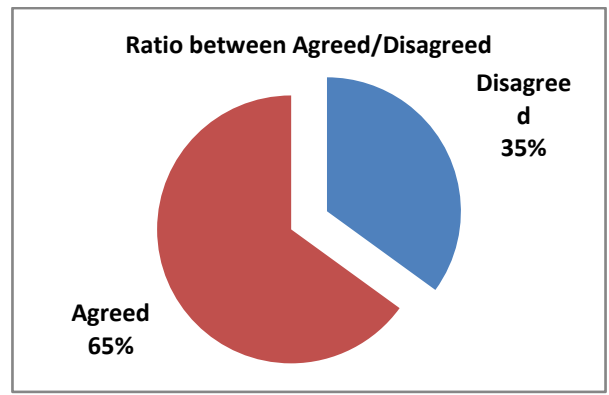

Figure 5.*Ratio between Agreed/Disagreed*